\documentclass{article} 

\usepackage{hyperref}
\usepackage{natbib}
\usepackage{hyperref}
\usepackage{algorithm}
\usepackage{algorithmic}
\usepackage{url}
\usepackage{comment}
\usepackage{amsmath, amssymb, natbib, graphicx, url}
\usepackage{caption}
\usepackage{subcaption}
\usepackage{amsmath}
\usepackage{bbm}
\usepackage{mathtools}
\usepackage{mathabx}
\usepackage{authblk}

\newtheorem{theorem}{Theorem}

\usepackage[margin=1.5in]{geometry}

\title{Private Hypothesis Testing for Social Sciences}

\author[1]{Ajinkya K Mulay}
\author[1, 2]{Sean Lane}
\author[1, 2]{Erin P. Hennes}

\affil[1]{\footnotesize Purdue University, West Lafayette, IN 47906, USA}
\affil[2]{\footnotesize University of Missouri, Columbia, MO 65211, USA}

\begin{document}
\maketitle
\begin{abstract}
  \noindent In conducting virtually any empirical study, whether experimental or observational, researchers routinely consider the amount of data that they need to or are able to collect in order to successfully find inferential support for their hypotheses. Sample size, and effect size, directly inform statistical power - the probability of observing a statistically significant effect of interest assuming such a true effect exists. However, as evidenced by the replication crisis and shift toward open science requirements in the social sciences, there is both misunderstanding and a lack the appropriate planning for determining optimal sample sizes that ensure adequate power. Careful planning ensures that the power remains high even under high measurement errors while keeping the Type 1 error (i.e., false positives) constrained. We study the impact of differential privacy on empirical studies and theoretically analyze the change in sample size required for adequate power due to the Gaussian mechanisms utilized to ensure non-identifiability of individuals with sensitive data. We show that extant differential privacy approaches reduce power, and thus the precision and replicability of hypothesized effects, especially in studies with limited sample size. Further, we provide an empirical method to improve the accuracy of private statistics with simple bootstrapping.
\end{abstract}

\maketitle
\section{Introduction}
One component of experimental planning involves designing the study in a way to optimize the statistical power or \textit{power} Cohen. \textit{power} is the probability that we observe a statistically significant empirical effect if such an effect exists in the population, and it is equal to $1 - \beta$, where $\beta$ is the Type 2 error (i.e., false negative). High statistical power is positively correlated with sample size since the measurement error is inversely proportional to the square root of the sample size. Therefore, we have the expression $SE \propto \frac{1}{\sqrt{N}}$ wherein $SE$ is the standard error and $N$ is the sample size. Next, we look at the current solutions that can help compute the appropriate sample size for empirical studies, spanning experimental and observational methods, to mitigate the impact of $SE$ on statistical inferences.

There are a variety of software solutions  \citep{lenth2001some} that provide the ability to compute optimal sample size in order to meet a certain \textit{power} threshold. Generally, in psychological studies, a common heuristic is to set this \textit{power} threshold to be $0.8$. However, we recognize that these solutions require strong assumptions on the point estimates of the p-value or effect size that are generally unknown and part of the overarching research agenda to identify. Failure of these assumptions is common and can convert a well-powered study into an underpowered one stemming from many factors, including random sampling variability \citep{gelman2006data}. \citep{card2020little} points out that a large number of Natural Language Processing (NLP) studies are underpowered and provides \textit{good} practices to counter these issues. \citep{gelman2006data} provides a universal solution for computing \textit{power} by simulating data from the same distribution as the data we intend to collect. Recently, \citep{mulay2021powergraph} provided a neural network-based method to predict \textit{power} over a large manifold of changing model coefficients and sample size. However, none of these works consider the impact of differential privacy on the corrected sample size. 

\citep{vu2009differential} is one of the first works to address this gap, and they provide the theoretical analysis to compute the corrected sample size ($N_{corrected}$) over Laplacian privacy for the z-test and the $\chi^{2}$ test. Alternatively, \citep{campbell2018differentially}, \citep{swanberg2019improved} demonstrates differentially private ANOVA with Laplacian noise. \citep{kifer2016new} demonstrates a generalized class of private $\chi^{2}$ tests. However, they do not provide analytical guarantees for the dataset size, and empirically they require a large dataset or even asymptotic sample sizes to reach the required level of statistical power. Practically, in the social sciences such data requirements are not possible, and so the utility of these methods for jointly optimizing privatization and robust inference is knowingly compromised. Here, we extend the contributions of \citep{vu2009differential}  and provide the following new results, 
\begin{itemize}
    \item Analytically compute the $N_{corrected}$ over Gaussian Noise for the z-test, t-test, f-test, and the $\chi^{2}$ test. Further, we provide insights into how a social scientist can use the usually known or assumed statistics in order to perform a private hypothesis test.
    \item Provide a universal empirical solution for computing statistics privately with bootstrapping. We could then privately compute these statistics from pilot studies for use elsewhere. Normally, this value is assumed to be given non-privately.
\end{itemize}
\section{Proposed Work}
\subsection{Differential Privacy} We follow the definition of \citep{dwork2014algorithmic}. For any two neighboring datasets $x, x{'} \subset \mathcal{S}$ differing over a single element, we say that the randomized mechanism $\mathcal{M}: \mathcal{S} \rightarrow \mathcal{R}$ is $(\epsilon, \delta)$ differentially private if, 
\begin{align*}
    \text{Pr}[\mathcal{M}(x) \in \mathcal{R}] \leq e^{\epsilon} \cdot \text{Pr}[\mathcal{M}(x^{'}) \in \mathcal{R}] + \delta
\end{align*}
For Laplacian or Exponential mechanisms, $\delta = 0$ while for Gaussian mechanism, $\delta$ is generally much smaller than $\frac{1}{N}$.

\subsection{Algorithms for introducing privacy}\label{section:algorithms} We assume that we can only access private statistics from the dataset in this work. Thus, even though we can query any value, we cannot see the actual dataset. Thus, we attempt to recreate the private statistic such that it is as close as possible to the original non-private statistic. Our strategy for identifying $N_{corrected}$ is to reduce the private and non-private statistic difference by tuning the sample size. Also, note that $N$ cannot be arbitrarily extended due to budget and logistical constraints in the number of available samples. Consider the linear model for the f-test,
\begin{align*}
    \hat{y} = \hat{\beta}_{1} x_{1} + \hat{\beta}_{2} x_{2} + ... + \hat{\beta}_{p} x_{p} + e
\end{align*}
where $\hat{y}$ is the predicted regression value, $x_{i} \in \mathbb{R}^{N}, \hat{\beta}_{i} \in \mathbb{R}$ are the predictors and the model coefficients respectively over $i \in [p]$ while $e$ is the measurement error given by $e_{i} \sim \mathcal{N}(0, \frac{\sigma^2}{N}), e \in \mathbb{R}^{N}$. Here $\sigma$ is the sample standard deviation and the sample size is $N$. Below we demonstrate the formulae of each statistic (denoted by $f(\cdot)$ and the privacy function is $\tilde{f}(\cdot) = f(\cdot) + $ Gaussian noise.
\begin{itemize}
    \item \textbf{z-test/t-test:} $f(x) = \frac{x - \mu}{\sigma}$, where $x$ is the observed mean and $\mu$ is the expected mean. $\tilde{f}(x) = f(x) + \mathcal{N}(0, (\frac{2cs\sqrt{q}}{N\epsilon})^2)$.
    \item \textbf{$\chi^{2}$ test:} $f(x, E) = \sum_{j=1}^{k} \frac{(x_{j} - E_{j})^2}{E_{j}}$ where $x$ are the observed frequencies and $E$ are the expected frequencies for the $k$ groups. $\tilde{f}(x, E) = f(x, E) + \mathcal{N}(0, (\mathcal{O}(\frac{\Delta_{j_{*}}}{\epsilon E_{j_{*}} \chi^{2}_{N}})^2))$, such that $j_{*} = \text{argmax}_{j} (x_{j} - E_{j}) \rightarrow \Delta_{j} = (x_{j} - E_{j}), \chi^{2}_{*}$ is the non-private statistic. We assume that the sample size is large enough so that the central limit theorem holds.
    \item \textbf{(partial) F-test:} $F_{statistic} =  (\frac{RSS_{R}}{RSS_{F}} - 1) \cdot (\frac{N-p}{p-r})$ where, $RSS$ is the residual sum of squares, $p$ is the dimension of the full model \textbf{(F)} and $r$ is the number of dimensions of the reduced model \textbf{(R)}. We can show using the properties of RSS and the Mediant inequality that the sensitivity of the $\frac{RSS_{R}}{RSS_{F}}$ will not change even after addition of a new sample given a maximum effect size given as $\Delta_{*}$ (deduced via a pilot or a user requirement). Thus, compute $(\frac{RSS_{R}}{RSS_{F}} + \mathcal{N}(0, \mathcal{O}(\Delta_{*}/\epsilon)- 1) \cdot N$ for a private F-test.
    \item \textbf{Sample Standard Deviation:} $\sigma = \sqrt{\frac{\sum_{i=1}^{N} (X_{ij} - \bar{X}_{j})^2}{N-1}}$ where $X_{i} \in \mathbb{R}^{p}, i \in [N], \bar{X}_{j} = \frac{1}{N} \sum_{i=1}^{N} X_{ij}$. $\tilde{\sigma} = \sigma + \mathcal{N}(0, (\frac{2cs\sqrt{q}}{\epsilon\sqrt{(N-1)}})^2) \rightarrow \frac{\tilde{\sigma}^{2}}{N} = \frac{\sigma^{2}}{N} + \mathcal{N}(0, \frac{4c^2qs^2}{\epsilon^2 (N-1)N})$ 
\end{itemize}

Here, each $X \in \mathbb{R}^{N \times p}, |X_{ij}| \leq s$ for $i \in [p], j \in [N]$, $q$ is the number of z-tests/t-tests done, $c = \sqrt{2 ln(1.25/\delta)}, c_{*} = c \cdot \sqrt{q}$. If we had used Laplacian noise, then instead of $\sqrt{q}$, we would have a factor of $q$ due to the use of L1 sensitivity instead of L2 in Laplacian noise.
 
\section{Computing the required sample size under privacy}\label{section:thm}
\begin{theorem}
For the z-test and the t-test, given the sample standard deviation $\sigma$, the sample size correction is given by $N_{correction} = \frac{1}{2}(1 + \sqrt{1 + \frac{k\gamma}{\epsilon^2 \sigma^4}})$, where for the z-test, $k = 16c^2qs^2/z_{*}, z_{*} = (z_{1-\alpha/2 + z_{1-\beta}})$ and for the t-test we can replace the $z_{*}$ by $t_{*} = (t_{1-\alpha/2 + t_{1-\beta}})$ as defined below.
\end{theorem}
We follow an analysis similar to \citep{vu2009differential} for the z-test or t-test. 
\begin{align*}
    \mathcal{H}_{0}: x = x_{0},  \mathcal{H}_{a}: x \neq x_{0}
\end{align*}

$\sigma, \gamma$ can be estimated or are available. $f(x) = \gamma = (x-x_{0})$. We add noise, $\mathcal{N}(0, \sigma_{*})$ where $\sigma_{*} = \frac{\sqrt{2 ln(1.25/\delta)q} \Delta_{2}f}{N \epsilon} = \frac{c \sqrt{q} \Delta_{2}f}{\epsilon}; \Delta_{2}f = \text{max}_{x, y, ||x-y||_{1} \leq 1} ||f(x) - f(y)||_{2} = \frac{2s}{N}$ where $|x_{i}| \leq s$. Therefore, $\sigma_{*} = \frac{2c\sqrt{q}s}{N\epsilon}$ and thus, we have two hypotheses and the corresponding equation for the critical sample size.
{\small
\begin{align*}
    \mathcal{H}_{0}:& \mathcal{N}(0, \frac{\sigma^2}{N} + \frac{4c^2q s^2}{N^2\epsilon^2})
    \\ \mathcal{H}_{a}:& \mathcal{N}(x-x_0, \frac{\sigma^2}{N} + \frac{4c^2 q s^2}{N^2\epsilon^2}) 
    z_{1-\alpha/2} (\frac{\sigma^2}{N} + \frac{4c^2q s^2}{N^2\epsilon^2}) =  \gamma - z_{1-\beta} (\frac{\sigma^2}{N} + \frac{4c^2 qs^2}{N^2\epsilon^2})
\end{align*}
}
For brevity, we leave the rest of the proof for the longer article and refer readers to \citep{vu2009differential}. Although this equation makes sense when $\sigma$ is known, the standard deviation is generally unknown. In this case, we could use the sample standard deviation based on a pilot study. Thus, we are now required to obtain a differentially private $\sigma$ variant based on the noise added to its computation. Although Algorithm~\ref{algorithm:PrivHistogram} (below) gives us a private estimate for the sample standard deviation, it does not fully answer the question of how the variability in noise affects the sample size correction. We leave this open problem for future work.

\begin{theorem}
For the chi-square test w.h.p we have that $N_{private} = N_{non-private} \cdot (1 + \mathcal{O}(\frac{\Delta_{j_{*}}}{\epsilon E_{j_{*}} \chi^{2}_{N}})) \leq N_{non-private} \cdot (1 + \mathcal{O}(\frac{1}{\epsilon \Delta_{j_{*}}}))$ (loose upper bound). The parameters are defined in \ref{section:algorithms}. Unlike previous literature, we do not add noise to the histogram but rather add it to the statistic itself with a data-dependent analysis, potentially introducing a much smaller noise level. In the case that our non-private analysis is `good,' we expect our private analysis to perform well due to the $\Delta_{j_{*}}$ (non-private) term being small. Note that in the case $E_{j}$ is unknown, we can easily replace it with the non-private estimate of $N$, which is assumed to be available or specified by the user. $\Delta_{j_{*}}$ is an effect size that the user can tune. Finally, $\chi^{2}_{N}$ can be estimated by pilot studies privately using Algorithm~\ref{algorithm:PrivHistogram}.
\end{theorem}
\begin{theorem}
For the F-test, w.h.p, the $N_{correction} = \frac{1}{1-\mathcal{O}(\frac{\Delta_{*}}{\epsilon})}$. Clearly, the $\epsilon$ value needs to be as large as possible to avoid a large correction and a smaller $\Delta_{*}$ helps reduce the correction as expected.
\end{theorem}

\section{Privatizing Statistics}\label{section:private-stats}

We look at a new algorithm based on the noisy histogram approach for calculating statistics as described in  Algorithm~\ref{algorithm:PrivHistogram}. The privacy analysis follows a usual noisy histogram algorithm with usual privacy aggregation. We assume that we have a multi-feature dataset and we are concerned with privatizing the column of the dataset with the assumption that the range of values of each feature is non-private. We notice that we will know the reasonable range of a feature in most cases. For example, for Diabetes patients, we can assume that we know the generalpopulation range of sugar values, or for heart disease patients, we know the blood pressure range. A significant advantage of this method is that we do not need to assume that our values belong to any specific distribution.

\begin{algorithm}
    \caption{Computing Private Statistics with Bootstrapping}
    \algsetup{linenosize=\tiny}
    \scriptsize 
    \label{algorithm:PrivHistogram}
    
    \begin{algorithmic}
    \STATE \textbf{Input}: Given data $X=\{x_{1}, x_{2}, ..., x_{N}\}$, $x_{i} \in \mathbb{R}$, Number of datasets $S$, Number of Histogram Bins $\mathcal{B}$
    \STATE $\mu^{*} \leftarrow$ Compute Global Statistic over $X$ (ex. mean)
    \STATE $X_{S} \leftarrow $ Create $S$ disjoint datasets each of size $\frac{|X|}{S}$ \COMMENT{assume $S$ completely divides $|X|$}
    \STATE $\mathcal{T}_{S} \leftarrow$ [ ]
    \FOR{disjoint dataset $d$ in $X_{S}$}
        \STATE $\mu_{d} \leftarrow$ Compute Statistic over $d$; Store $\mu_{d}$ in $\mathcal{T}_{S}$
    \ENDFOR
    \STATE $\mathcal{H}, \mathcal{M} \leftarrow$ Create Histogram from $\mathcal{T}_{S}$ with $\mathcal{B}$ bins \COMMENT{$\mathcal{H}$ (heights of histograms), $\mathcal{M}$ (means of each bin)}
    \STATE For each bin's height $\hat{\mathcal{H}_{i}} \leftarrow \mathcal{H}_{i} + \mathcal{N}(0, (\frac{c \cdot h}{\epsilon})^{2})$ \COMMENT{sensitivity is the number of histograms $h$}
    \STATE $\hat{\mu} \leftarrow \frac{\sum_{j=1}^{\mathcal{B}} \hat{\mathcal{H}_{j}} \cdot \mathcal{M}_{j}}{ \sum_{j=1}^{\mathcal{B}} \hat{\mathcal{H}_{j}}}$ \COMMENT{We assume that range of each bin is non-private}
    \STATE Privacy cost $\epsilon_{total} \leftarrow \epsilon \cdot k$ ($k$ is the maximum number of times any data-point is sampled)
 \end{algorithmic}
\end{algorithm}

\begin{figure}[htb!]
     \centering
     \begin{subfigure}[htb!]{0.4\textwidth}
        \includegraphics[width=\textwidth, scale=0.3]{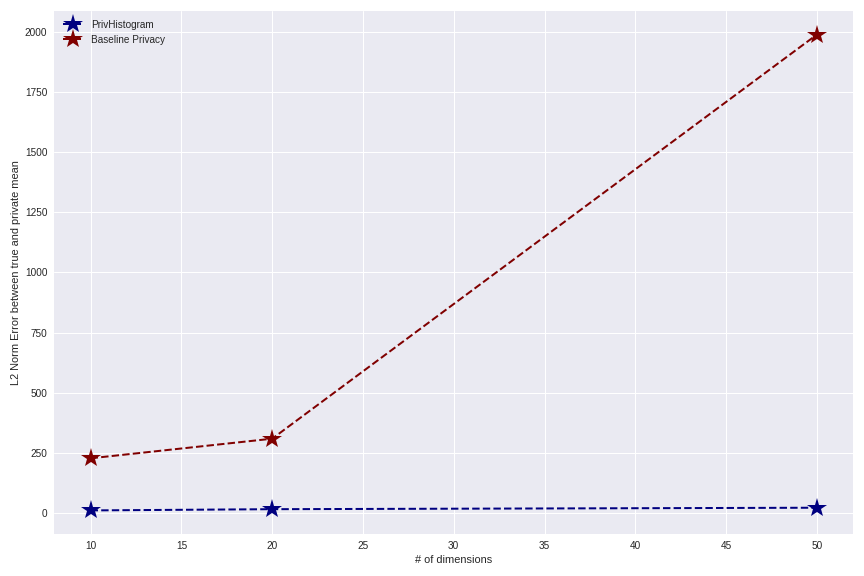}
        \caption{Private Mean}
        \label{fig:y equals x}
    \end{subfigure}
    \hfill
    \begin{subfigure}[htb!]{0.4\textwidth}
        \includegraphics[width=\textwidth, scale=0.3]{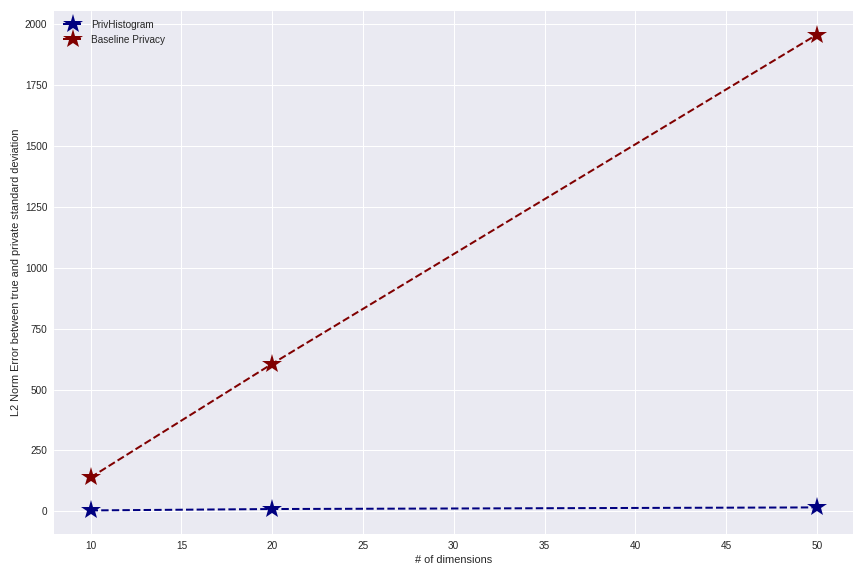}
        \caption{Private Standard Deviation}
        \label{}
    \end{subfigure}
    \caption{\textbf{PrivHistogram:} Comparison of errors over private mean and standard deviation derived using \textit{PrivHistogram} and \textit{vanilla} differential privacy. Sensitivity computed after looking at true data. Here, $\epsilon_{total} = 1.0$}
    \label{fig:three graphs}
\end{figure}

\section*{Future Work} The article provides sample size corrections with Gaussian noise along with a general method to compute private statistics. Our future work includes providing, (1) the in-depth theoretical analysis of the sensitivity and Algorithm \ref{algorithm:PrivHistogram}, and (2) empirical results for items described in sections~\ref{section:algorithms}, \ref{section:thm} and \ref{section:private-stats}.

\bibliography{ref}
\bibliographystyle{abbrvnat}

\appendix
\end{document}